\documentclass[fleqn,usenatbib]{mnras}

\usepackage{newtxtext,newtxmath}

\usepackage[T1]{fontenc}
\usepackage{ae,aecompl}
\usepackage{soul}

\usepackage{graphicx}	
\usepackage{amsmath}	
\usepackage{amssymb}	
\usepackage[usenames]{xcolor}
\usepackage{natbib}
\usepackage{hyperref}
\usepackage{xfrac}

\def\equationautorefname~#1\null{equation~(#1)}
\newcommand{\autorefp}[1]{%
  \begingroup%
  \def\equationautorefname~##1\null{equation~##1\null}%
  \autoref{#1}%
  \endgroup%
}

\DeclareMathAlphabet{\mathpzc}{OT1}{pzc}{m}{it}\definecolor{purple}{RGB}{160,32,240}
\newcommand{\Mh}{M_{\mathrm{h}}}
\newcommand{\Msun}{\,M_{\odot}}

\newcommand{\Mgc}{M_{\mathrm{GC}}}
\newcommand{\Mgcobs}{M_{\mathrm{GC,obs}}}

\newcommand{\Mstar}{M_{\star}}

\newcommand{\ttid}{t_{\mathrm{tid}}}

\newcommand{\feh}{\mathrm{[Fe/H]}}
\newcommand{\Mc}{M_c}

\newcommand{\Mtot}{M_{\mathrm{tot}}}
\newcommand{\fred}{f_{\mathrm{red}}}

\title[Origins of GC Scaling Relations]{Origins of scaling relations of globular cluster systems}

\author[Choksi \& Gnedin]{
Nick Choksi$^{1}$\thanks{E-mail: nchoksi@berkeley.edu} and
Oleg Y. Gnedin$^{2}$
\\
$^{1}$Department of Astronomy, University of California at Berkeley, Berkeley, CA, 94720, USA\\
$^{2}$Department of Astronomy, University of Michigan, Ann Arbor, MI, 48109, USA
}

\date{Released \today}

\pubyear{2019}

\begin{document}
\label{firstpage}
\pagerange{\pageref{firstpage}--\pageref{lastpage}}
\maketitle

\begin{abstract}
Globular cluster (GC) systems demonstrate tight scaling relations with the properties of their host galaxies. In previous work, we developed an analytic model for GC formation in a cosmological context and showed that it matches nearly all of the observed scaling relations across 4 orders of magnitude in host galaxy mass. Motivated by the success of this model, we investigate in detail the physical origins and evolution of these scaling relations. The ratio of the combined mass in GCs $\Mgc$ to the host dark matter halo mass $\Mh$ is nearly constant at all redshifts, but its normalization evolves by a factor of $\sim$10 from birth to $z=0$. The relation is steeper than linear at halo masses $\Mh \lesssim 10^{11.5}\Msun$, primarily due to non-linearity in the stellar mass-halo mass relation. The near constancy of the ratio $\Mgc/\Mh$, combined with the shape of the stellar mass-halo mass relation, sets the characteristic $U-$shape of the GC specific frequency as a function of host galaxy mass. The contribution of accreted satellite galaxies to the buildup of GC systems is a strong function of the host galaxy mass, ranging from $\approx$0\% at $\Mh \approx 10^{11} \Msun$ to 80\% at $\Mh \approx 10^{15} \Msun$. The metal-poor clusters are significantly more likely to form ex-situ relative to the metal-rich clusters, but a substantial fraction of metal-poor clusters still form in-situ in lower mass galaxies. Similarly, the fraction of red clusters increases from $\approx 10\%$ at $\Mh = 10^{11} \Msun$ to $\approx 60\%$ at $\Mh \approx 10^{13} \Msun$, and flattens at higher $\Mh$. Clusters formation occurs essentially continuously at high redshift, while at low redshift galactic mergers become increasingly important for cluster formation.
\end{abstract}



\section{Introduction}
\label{sec:Intro}

The combination of their high masses, small sizes, and old ages means that globular clusters (GCs) in the local universe can be used as fossils of the extreme, high-redshift environments in which they formed. Although the upcoming James Webb Space Telescope and future thirty-meter class telescopes are poised to bring direct observations of GC \textit{formation} at high-redshift \citep{renzini_2017, boylan-kolchin_2017, zick_etal_2018, vanzella_etal_2017, vanzella_etal_2019}, current observations of GCs are limited to within 200 Mpc \citep{harris_etal14_bcg1}. Inside this range, however, there is now a wealth of information on GC systems \citep[e.g.,][]{huchra_etal_1991, harris96, beasley_etal_2000a, beasley_etal_2000b, peng_etal06, leaman_etal13, brodie_etal_2014, harris_etal14_bcg1, harris_etal16_bcg2, harris_etal17_bcg3, forbes_etal18}. Despite the diversity of systems studied, observations have revealed that the ensemble properties of GC systems consistently correlate with the properties of their host galaxies. Below, we briefly summarize the main properties of GC systems relevant to this work.

The combined mass in GCs $\Mgc$ has been shown to scale almost linearly with the total host galaxy mass $\Mh$ (including the dark matter halo) over at least 4 orders of magnitude, with a normalization \citep{spitler_forbes09, hudson_etal_2014, harris_etal17, forbes_etal_2017, forbes_etal_2018}:
\begin{equation}
  \Mgc \approx 3.4 \times 10^{-5}\Mh. \nonumber
\end{equation} 
On the other hand, the specific frequency of GCs (the number of GCs divided by the galaxy stellar mass or luminosity) is highly non-linear and shows a characteristic $U$-shape, which is minimized for galaxies of stellar mass $\Mstar \sim 10^{10} \Msun$. Moreover, within an individual galaxy, the specific frequency also decreases with metallicity \citep{harris_harris02, beasley_etal_2008, lamers_etal_2017}.

In contrast to the relatively universal GC mass function, the metallicity distribution function (MDF) of GC systems varies greatly from galaxy to galaxy. Because of the difficulty in acquiring high-resolution spectroscopy of extragalactic GCs, these MDFs are typically derived from integrated photometry combined with an empirically calibrated colour-metallicity relation \citep{peng_etal06, harris_etal06, usher_etal_2012}. The mean and dispersion of GC system MDFs have also been found to scale with the mass of the host galaxy over a large range in galaxy mass \citep{peng_etal06}. Qualitatively, many galaxies have clearly multi-modal MDF \citep[e.g.,][]{gebhardt_kissler-patig99, peng_etal06, brodie_strader06, brodie_etal_2012}, but the most massive galaxies typically show broad, single-peaked distributions \citep{harris_etal14_bcg1, harris_etal16_bcg2, harris_etal17_bcg3}.

The discovery of multi-modality in GC system MDF has historically led to the division of GCs into the so-called metal-poor ``blue'' and metal-rich ``red'' subpopulations, with a typical dividing line at $\feh \approx -1$. These two subpopulations also differ in their spatial and kinematic properties. The blue GCs are typically on radially extended, eccentric orbits supported by random motions, while the red GCs follow more closely the field star distribution and kinematics \citep{zinn85,strader_etal_2011, forbes_etal_2011, pota_etal_2013, durrell_etal_2014, schuberth_etal10}. Motivated by these observations, many authors have since suggested that the blue GCs formed ex-situ, in now-disrupted satellite galaxies, while the red GCs formed primarily in-situ \citep[e.g.,][]{cote_etal98, katz_ricotti_2014}. On the theoretical side, recent efforts have sought to build unified models of GC formation as a natural outcome of normal star formation occurring at the highest densities and pressures \cite[][]{kravtsov_gnedin05, muratov_gnedin10,  tonini2013, li_gnedin14, kruijssen2015, li_etal_sim1, pfeffer_etal_2018, choksi_etal_2018}, rather than invoking distinct formation physics for each of the two subpopulations \citep[as proposed by, e.g., ][]{forbes_etal97, cote_etal98, beasley_etal02, strader_etal05, griffen_etal10}. 

Observations of ubiquitous young massive star clusters by the Hubble Space Telescope in nearby interacting and starbursting galaxies, such as the spectacular Antennae galaxies \cite[e.g.,][]{whitmore_etal93}, have solidified the idea that mergers of gas-rich galaxies may trigger formation of star clusters destined to evolve into globular clusters. Galaxy mergers can pressurize the ISM and provide tidal torques that ultimately drive large-scale inflows of cold gas \citep{ashman_zepf92, kravtsov_gnedin05, bournaud_etal08, renaud_etal_2015, li_etal_sim1}, increasing the bound cluster formation efficiency \citep{kruijssen12}. 

In \cite{choksi_etal_2018}, we presented an analytic model for the formation and evolution of GC systems along these lines. In this model, periods of rapid accretion onto dark matter halos are assumed to trigger cluster formation. The properties of the cluster population that form are set by the properties of the host galaxy, which are in turn set by empirical scaling relations derived from the observational literature. We found that GCs form over a wide range of redshifts and host galaxy masses, but the majority of cluster formation still occurs before the peak of field star formation. Furthermore, we demonstrated that this simple model simultaneously matches nearly all of the scaling relations obeyed by GC systems.

In \cite{choksi_gnedin_19} we improved on the physical realism of this model by updating the adopted cluster initial mass function from a pure power-law to a \citet{schechter_1976} function $dN/dM \propto M^{-2}e^{-M/\Mc}$ which exponentially truncates the formation of the most massive clusters. This change was motivated by numerous observations of the initial mass function of young clusters in nearby galaxies \citep[e.g.,][]{gieles_etal_2006a, larsen09, bastian_2008, adamo_etal_2015, johnson_etal_2017} and by galaxy formation simulations that model cluster formation \citep{li_etal_sim1, li_etal_sim2}. After experimenting with various values of the cutoff mass, we determined that $\Mc \approx 10^{6.5}-10^{7}\Msun$ produced results that were most consistent with present-day GC scaling relations. Lower values of $\Mc$ prevent the formation of very massive GCs found in massive elliptical galaxies \citep{harris_etal14_bcg1}, while higher values are disfavoured by modeling the high-mass end of the present-day GC mass function \citep{johnson_etal_2017}. \autoref{sec:methodology} provides more detail on the model setup.

In this work, we investigate the origin of the scaling relations presented in \cite{choksi_etal_2018}. We begin in \autoref{sec:mgc_mh} by examining the evolution over time of the GC system mass - halo mass ($\Mgc-\Mh$) relation. We also discuss the origins of non-linearity in the relation for dwarf galaxies and the implications for the specific frequencies of GCs. In \autoref{sec:satellites} we investigate the contribution of accreted satellite galaxies and their GC systems to the properties of present-day GC systems. In \autoref{sec:mergers} we discuss the role of galaxy mergers in GC formation. Finally, in \autoref{sec:discussion} we discuss the implications of our results and summarize our  conclusions.

\section{Methodology} 
\label{sec:methodology}

Below, we briefly summarize the main components of our cluster formation model. For further details and justification regarding the choices of equations and parameters, we refer readers to \cite{choksi_etal_2018} and \cite{choksi_gnedin_19}.

\subsection{Summary of cluster formation model}

We trigger cluster formation when the specific mass accretion rate $R_{\rm m}$ onto a dark matter halo exceeds a threshold value $p_3$ between consecutive outputs of our adopted collisionless cosmological simulation. Specifically, for a halo of mass $M_{\rm h,2}$ at time $t_2$ and its progenitor of mass $M_{\rm h,1}$ at time $t_1$, we calculate the specific mass accretion rate as:
\begin{equation}
  R_{\rm m} \equiv \frac{M_{\rm h,2} - M_{\rm h,1}}{t_2 - t_1}. \frac{1}{M_{\rm h,1}},
  \label{eqn:rm}
\end{equation}
and trigger cluster formation for $R_{\rm m} > p_3$. When cluster formation is triggered, we form a population of clusters of combined mass $\Mtot$:
\begin{equation}
    \Mtot = 1.8 \times 10^{-4}p_2 M_g,
    \label{eqn:mtot}
\end{equation}
where $M_g$ is the cold gas mass in the galaxy and the normalization of \autoref{eqn:mtot} is motivated by the cosmological hydrodynamic simulations of \cite{kravtsov_gnedin05}. The values of $p_2$ and $p_3$ are taken to be free parameters\footnote{The current form of our model has only two adjustable parameters, but we keep the labels $p_2$ and $p_3$ for consistency with previous versions of the model.}, whose values are constrained using a wide array of observational data on GC system metallicities and masses \citep{harris96, cote_etal06, peng_etal06, harris_etal14_bcg1, harris_etal16_bcg2, harris_etal17_bcg3}. 

The cold gas fraction is parameterized as a function of the stellar mass $\Mstar$ and redshift $z$ as: 
\begin{equation}
  \frac{M_g}{\Mstar}(\Mstar, z) = 0.35 \times 3^{2.7} \, \left(\frac{\Mstar}{10^9\Msun}\right)^{-n_m(\Mstar)} \ \left(\frac{1+z}{3}\right)^{n_z(z)},
  \label{eqn:fg}
\end{equation}
where $n_z$ and $n_m$ are given by:
\begin{align}
  n_z &= 1.4 \;\mathrm{for}\; z > 2, \;\mathrm{and}\; n_z = 2.7 \;\mathrm{for}\; z < 2, \nonumber \\
    n_m &= 0.33 \;\mathrm{for}\; \Mstar > 10^{9}\Msun, \;\mathrm{and}\; n_m = 0.19 \;\mathrm{for}\; \Mstar < 10^{9}\Msun.
  \nonumber
\end{align}
The host galaxy stellar mass is increased self-consistently using a modified version of the semi-empirical stellar mass-halo mass relation derived from forward modeling by \cite{behroozi_etal_2013_main}, and extrapolated at high-redshift ($z \gtrsim 8$). The stellar mass also sets the galaxy metallicity via an empirical galaxy mass-metallicity relation:
\begin{equation}
  \feh =  \log_{10}\left[\left(\frac{\Mstar}{10^{10.5}\Msun}\right)^{0.35} (1+z)^{-0.9}\right].
  \label{eq:mmr}
\end{equation}
Individual cluster metallicities are set by the metallicity of the host galaxy in which they form, with an additional scatter of 0.3 dex. Clusters are drawn from a cluster initial mass function of the form:
\begin{equation}
    \frac{dN}{dM} = M_0M^{-2}e^{-M/\Mc},
\end{equation}
where $M_0$ is an overall normalization factor and $\Mc$ is the characteristic truncation mass. Our method for sampling the cluster initial mass function is the ``optimal sampling'' method of \cite{schulz_etal_2015}.

In \citet{choksi_gnedin_19} we analyzed the effects of different values of $\Mc$ on GC system scaling relations. However, because it is not the focus of this work, here we present results from only one model of fixed $\Mc = 10^{7} \Msun$. This value of $\Mc$ is consistent with theoretical expectations and inferences from observations of the local GC mass functions, and also robustly reproduces the various GC system scaling relations \citep{reina-campos_kruijssen_2017, jordan_etal_2007, johnson_etal_2017, choksi_gnedin_19}. All results presented in this work are qualitatively robust to variations in $\Mc$. For the $\Mc = 10^{7} \Msun$ model we adopt in this work, the best-fit values of the free parameters are $p_2=8.8$, $p_3 = 0.58 \rm \,Gyr^{-1}$.

GCs lose mass gradually due to two-body relaxation and tidal stripping. Because our model contains no spatial information on cluster orbits, we apply a spatially-averaged dynamical disruption prescription following \cite{gnedin_etal14}:
\begin{equation}
    \frac{dM}{dt} = -\frac{M}{\ttid},
    \label{eqn:dmdt}
\end{equation}
where the disruption timescale $\ttid$ was calibrated using direct $N-$body simulations \citep{gieles_baumgardt08}:
\begin{equation}
    \ttid(M) = 5 \, \mathrm{Gyr} \left(\frac{M(t)}{2 \times 10^5 \Msun}\right)^{2/3}\left(\frac{P}{0.5}\right),
    \label{eqn:ttid}
\end{equation}
where $P$ is a normalized period of rotation around the galactic center. In \cite{choksi_gnedin_19} we showed that adopting a constant value of $P=0.5$ reproduces well the present-day GC mass function. Integrating \autoref{eqn:dmdt} yields the mass as a function of time $t$ since cluster birth:
\begin{align}
  M(t) = M(t=0)\left[ 1 - \frac{2}{3}\frac{t}{t_{\mathrm{tid}}(t=0)} \right]^{3/2}\times f_{\rm lost, se}(t),
\end{align}
where we apply a fractional mass loss due to stellar evolution $f_{\rm lost, se}(t)$ as calculated by \cite{prieto_gnedin08}.

\section{The globular cluster system mass - halo mass relation}
\label{sec:mgc_mh}

In this section we investigate the origin and evolution of the relationship between the combined mass in GCs and the host halo mass. Throughout this section and next, we distinguish between the central galaxy and its satellites using the ``main progenitor branch'' (MPB) tag of the adopted dark matter merger trees. The MPB is defined as the branch along the merger tree with the largest \textit{integrated} mass history \citep[for details, see][]{de-lucia_blaizot_2007}. It can be thought of as a ``current'' halo mass at each redshift. Merger trees are taken from the collisionless run of the \textit{Illustris} simulation and constructed using the \textsc{sublink} algorithm \citep{springel_etal_2001, vogelsberger_etal_2014}.

\begin{figure}
\includegraphics[width=\columnwidth]{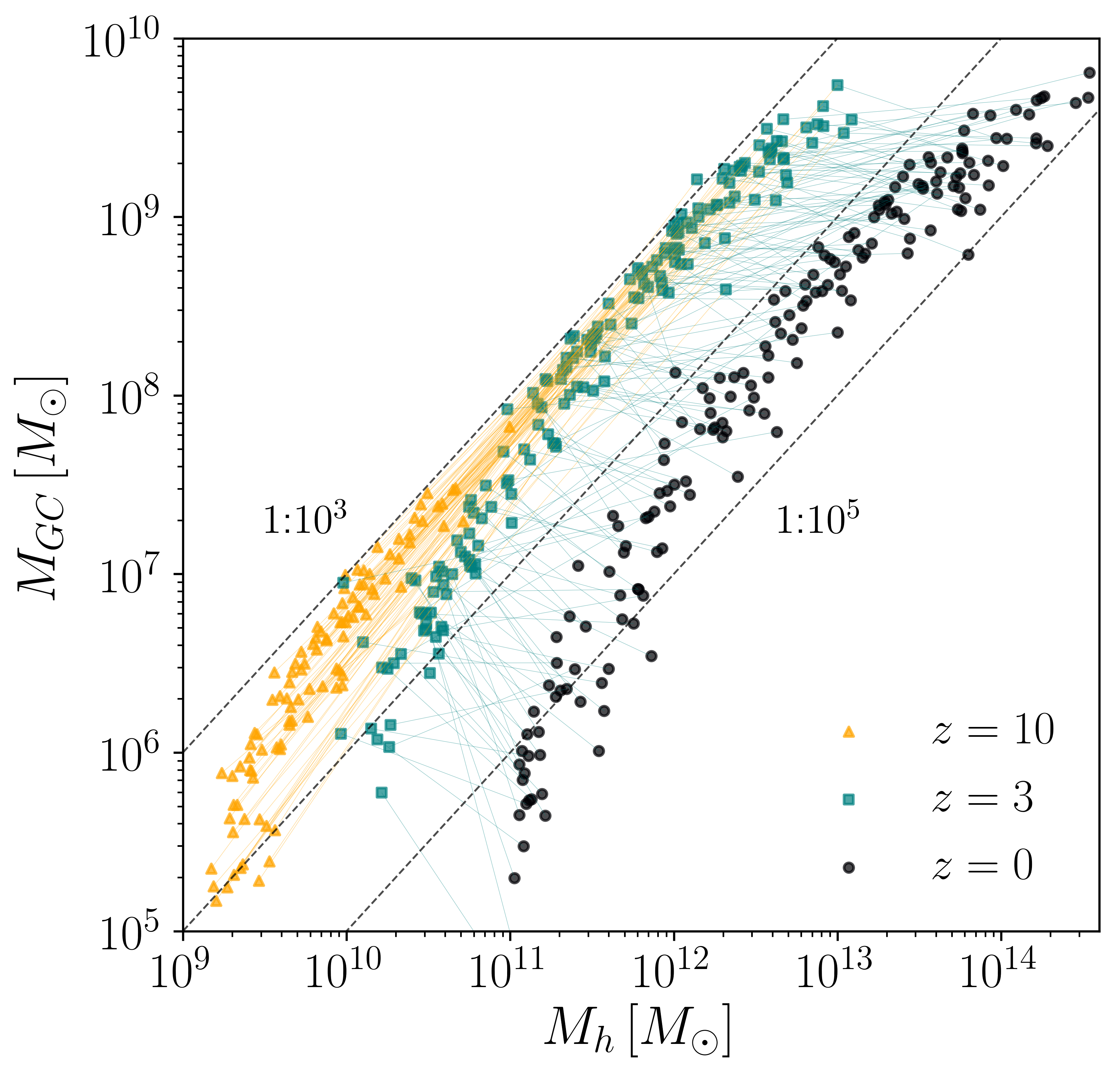}
\vspace{-5mm}
\caption{Evolution of the $\Mgc-\Mh$ relation. Each point shown represents the main progenitor branch (MPB) halo and the bound GC mass within it at a given redshift, including disruption and stellar evolution. Thin lines connect the same halo across epochs. Halos are only identified by the halo finder once they have sufficient number of particles, and so there are fewer points at higher redshift as some halos have not yet been identified. For reference, the three dashed lines show normalization of $\Mgc/\Mh = 10^{-5}, 10^{-4}$, and $10^{-3}$. }
  \label{fig:mgcz3}
\end{figure}

\autoref{fig:mgcz3} shows the $\Mgc-\Mh$ relation at three distinct epochs, $z=10$, 3, and 0. We include in the calculation of $\Mgc$ only the clusters that have already been accreted onto the MPB, so that $\Mgc$ represents the observable mass in clusters in the galaxy at a given epoch, including the effect of disruption and stellar evolution. $\Mh$ is the mass of the MPB halo of each merger tree. We find that the nearly linear relationship between $\Mgc$ and $\Mh$ holds across redshifts, but with a normalization $\Mgc/\Mh$ that decreases monotonically with cosmic time.

When cluster formation is triggered, we form a population of clusters of combined mass $\Mtot \sim 1.6\times 10^{-3}M_g$. At $z \sim 10$, near the beginning of cluster formation, most galaxies are very gas-rich, and $M_g \sim f_b \Mh$. Each cluster formation event then produces a mass in clusters $\Mtot \sim 3\times 10^{-4}\Mh$. There are several outputs of the \textit{Illustris} simulation at $z>10$, which may combine the young cluster populations to reach $\Mgc \sim 10^{-3}\Mh$. This is the upper envelope of the normalization seen at high redshifts in \autoref{fig:mgcz3}.

At high redshift, continuous cluster formation and accretion of external GC systems during galaxy mergers, combined with concurrent growth of the host halo, causes points to move diagonally in the $\Mgc-\Mh$ plane. The normalization decreases only by a factor of $\sim2$ from $z=10$ to $z=3$.

On the other hand, we find that between $z=3$ and $z=0$ the normalization decreases by a much larger factor of $\sim10$: from $\Mgc \approx 3\times 10^{-4} \Mh$ at $z=3$ to $\Mgc \approx 3\times 10^{-5} \Mh$ at $z=0$. Over this redshift range, the GC system mass can decrease because of dynamical disruption, or increase because of late cluster formation, but on average it remains nearly constant. In contrast, the halo mass only increases with time and this dominates the evolution of the normalization at $z \lesssim 3$ and causes the points in Fig.~\ref{fig:mgcz3} to move rightward in the $\Mgc-\Mh$ plane. Furthermore, because the logarithmic accretion rates onto dark matter halos depend only weakly on halo mass, $\dot{\Mh}/\Mh \propto \Mh^{0.13}$ \citep{mcbride_etal_2009}, the points shift rightward about the same amount and therefore the shape of the $\Mgc-\Mh$ relation is preserved until $z=0$.

However, our model also predicts deviation from a purely linear relation. At halo masses below $\Mh \sim 10^{11.5} \Msun$ the predicted $\Mgc-\Mh$ scaling falls off faster than linearly. Moreover, this behaviour is already present at $z \sim 10$, meaning that it is imprinted at birth and is not due to how the GC system evolves. The shape of the relation between the cold gas mass and halo mass therefore is the dominant factor in setting the model's non-linearity. The $M_g - \Mh$ relation is set indirectly, via the $M_g - \Mstar$ relation in \autorefp{eqn:fg} and the $\Mstar - \Mh$ relation of \cite{behroozi_etal_2013_main}. The latter of these is highly nonlinear and dominates the nonlinear behaviour seen in the model. Therefore, the observed non-linearity in Fig.~\ref{fig:mgcz3} can be used to place a constraint on the shape of the stellar mass-halo mass relation at high redshift.

To quantify the degree of non-linearity, we perform a linear fit to the model $\Mgc-\Mh$ relation at each redshift and calculate the median deviation (in perpendicular log-mass distance) of the model points from the best-fit line. At $z > 3$, this deviation remains at small values 0.13-0.14 dex, barely distinguishable from the linear relation. By $z=0$ the median deviation increases to 0.19 dex, and the relation can be considered mildly non-linear.

At the largest galaxy masses, the $\Mstar - \Mh$ relation \citep{behroozi_etal_2013_main} is significantly more non-linear than the $\Mgc-\Mh$ relation. Giant early-type galaxies are believed to assemble by consuming a large number of satellite galaxies, thus combining both their field stellar populations and GC systems. Why does this process result in visibly different scaling relations for $\Mstar$ and $\Mgc$? We return to this point in next section where we investigate the contribution of satellite galaxies to the overall GC system.

Finally, although the non-linearity is already present at birth, we note that the lower average cluster masses in low-mass galaxies \citep{choksi_etal_2018}, combined with the fact that low-mass clusters have larger fractional mass loss by \autoref{eqn:ttid}, may exacerbate the steeper than linear fall-off in dwarf galaxies. We plan to investigate the GC systems of dwarf galaxies in more detail in future work.

\begin{figure}
\includegraphics[width=\columnwidth]{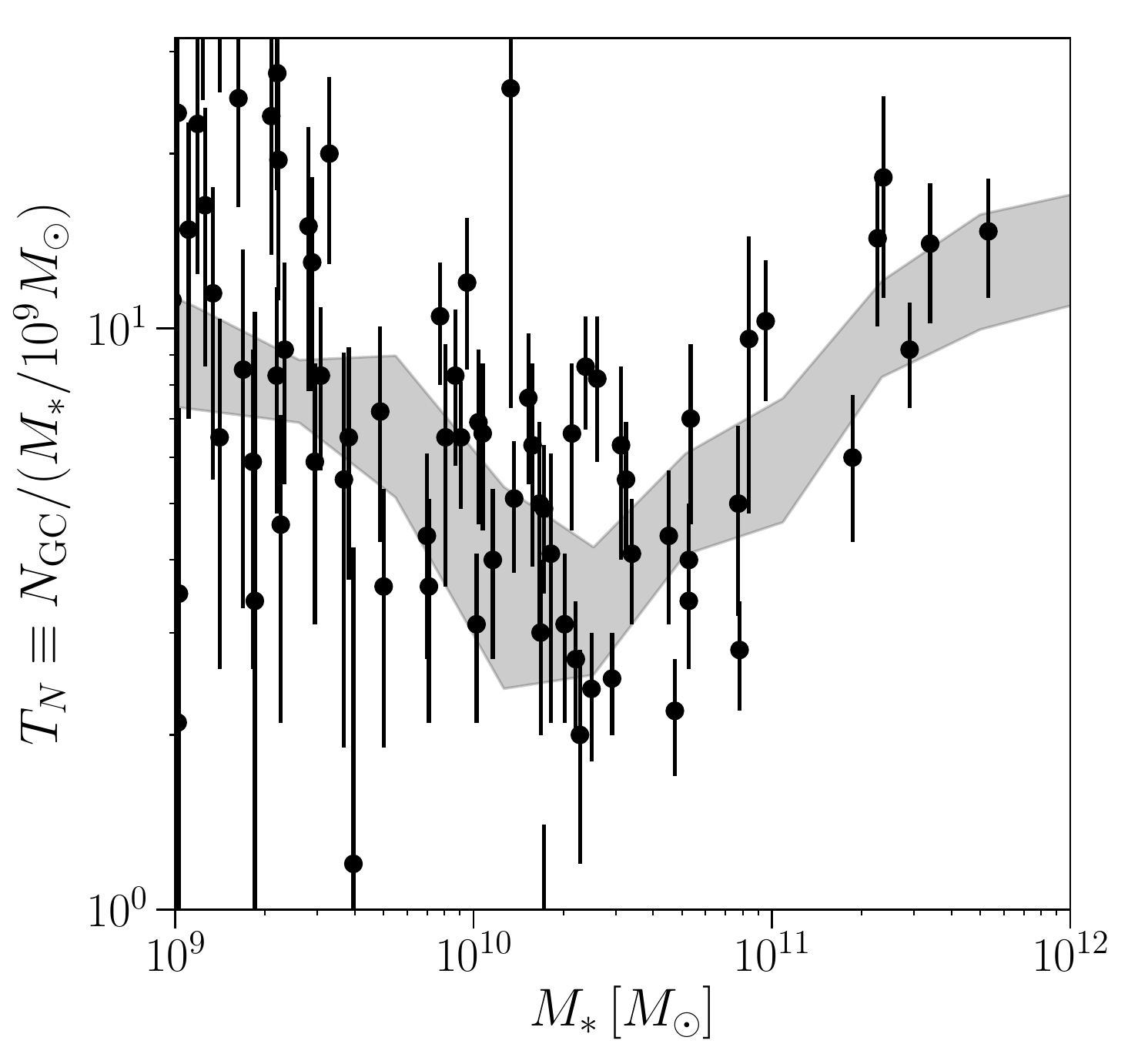}
\vspace{-5mm}
\caption{Specific frequency as a function of host galaxy mass. The model prediction is shown as the gray shaded region (interquartile range), with data from the Virgo Cluster Survey \citep{peng_etal08} shown as black points. The characteristic U-shape is set by the near constancy of the $\Mgc/\Mh$ ratio and the peaked shape of the $\Mstar/\Mh$ ratio.}
  \label{fig:tn}
\end{figure}

A commonly used statistic in studying GC systems is the specific frequency $T_N \equiv N_{\rm GC}/(\Mstar/10^9 \Msun)$. This specific frequency is proportional to
$$
   T_N \propto \frac{\Mgc}{\Mh}\frac{\Mh}{\Mstar}.
$$
The first ratio in the above expression, $\Mgc/\Mh$, is nearly constant. The ratio $\Mh/\Mstar$, however, is a strong function of $\Mstar$ and is minimized for $\Mstar \approx 10^{10.2} \Msun$ \citep[e.g.,][]{behroozi_etal_2013_main}. Thus, in our model the $U-$shape of the specific frequency as a function of galaxy mass is predominantly set by the shape of the stellar-mass halo mass relation (see also \citealt{el-badry_etal_2019}). \autoref{fig:tn} shows that our model prediction for $T_N$ is consistent with data from the Virgo Cluster Survey (as presented in \citealt{peng_etal08}). This is expected given that the model matches the $\Mgc-\Mh$ relation. At low halo masses, the ratio $\Mgc/\Mh$ is no longer constant and its variation with host mass will begin to affect the shape of the $T_N$ relation.

\section{Effects of satellite systems}
\label{sec:satellites}

In this section we examine the impact of accreted satellite systems on the buildup of present-day GC systems. As discussed in \autoref{sec:mgc_mh}, we identify the central galaxy and its satellites at each redshift using the ``main progenitor branch'' tag of the adopted dark matter merger trees.

\begin{figure}
\includegraphics[width=\columnwidth]{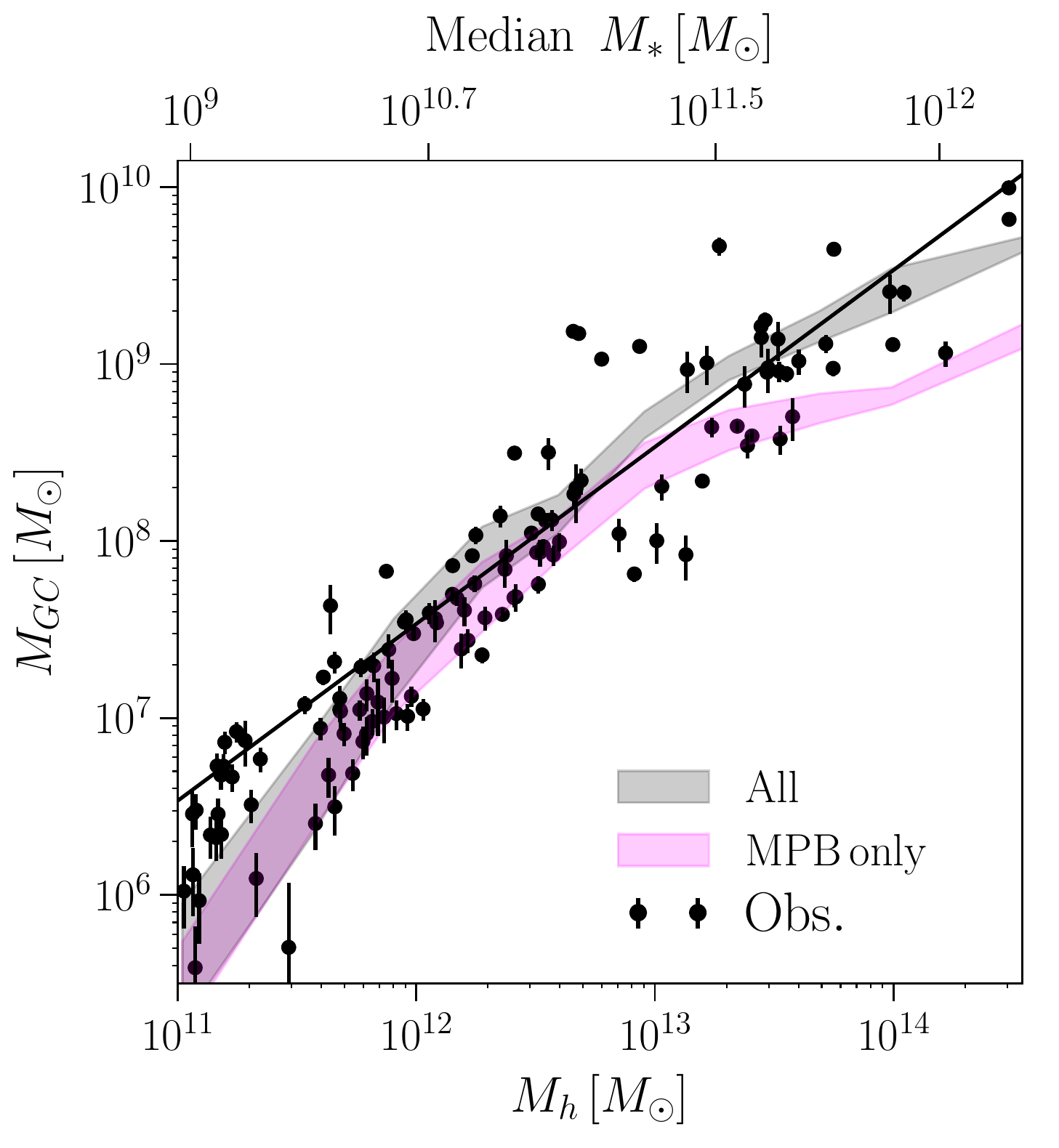}
\vspace{-5mm}
\caption{Effect of excluding satellites on the $\Mgc-\Mh$ relation. The grey shaded region shows the interquartile range of the fiducial model result. The magenta shaded region shows the fiducial model result, but excluding the contribution of any GCs formed ex-situ.}
  \label{fig:m_mpb}
\end{figure}

\autoref{fig:m_mpb} illustrates the effect of satellite GC systems on the $\Mgc-\Mh$ relation. It shows the fiducial model prediction with and without contribution of GCs formed in satellite systems. At low halo masses, the two cases are indistinguishable. However, the difference becomes noticeable in the most massive (group or cluster sized) halos with $\Mh \gtrsim 10^{13}\Msun$. Without the contribution of accreted GCs, the GC system mass is too low by up to 0.3~dex.

\begin{figure}
\includegraphics[width=\columnwidth]{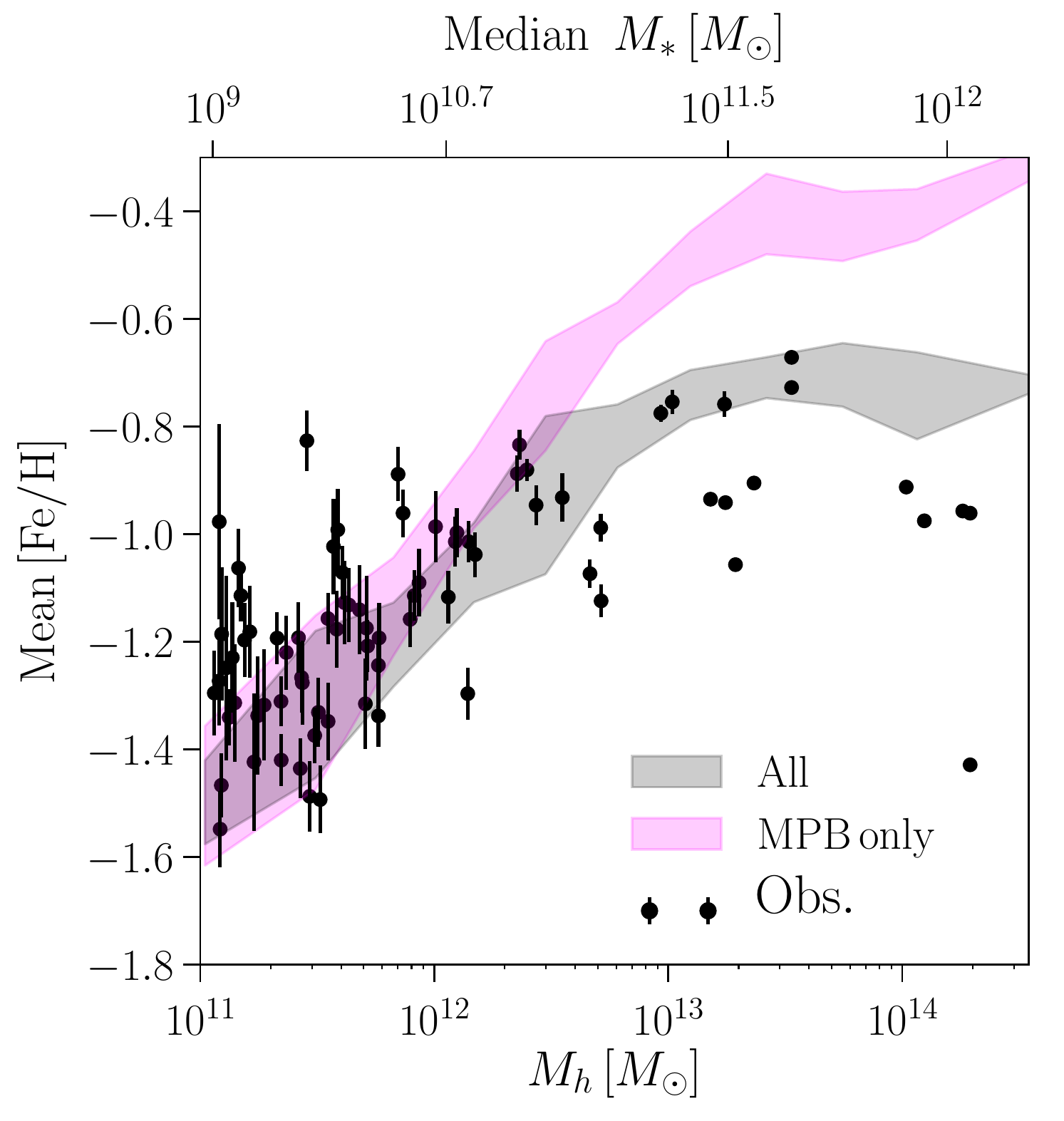}
\vspace{-5mm}
\caption{Effect of accreted systems on the mean metallicity of GC systems as a function of host halo mass. Legend as in \autoref{fig:m_mpb}.}
  \label{fig:mean_mpb}
\end{figure}

\begin{figure}
\includegraphics[width=\columnwidth]{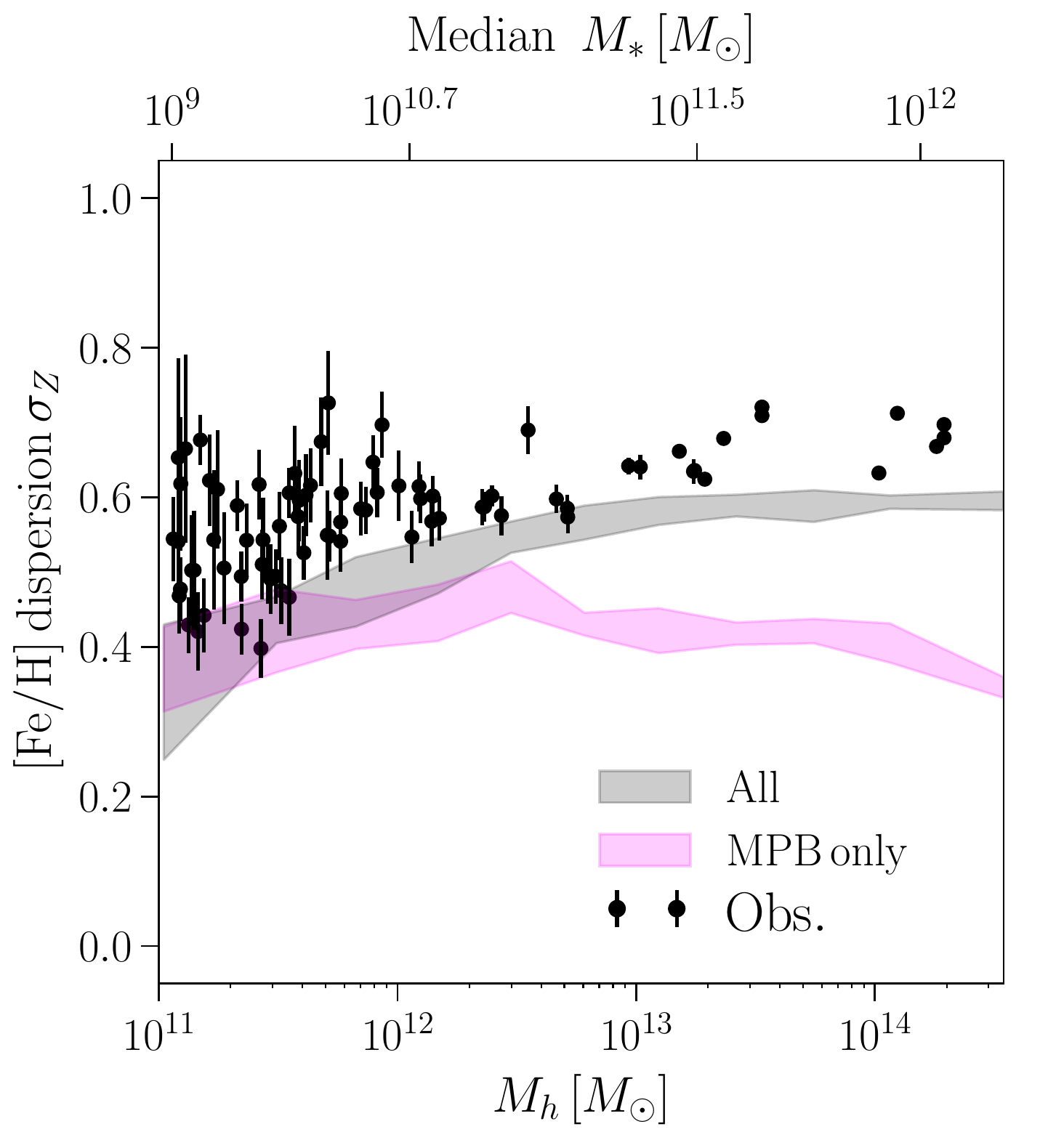}
\vspace{-5mm}
\caption{Effect of accreted systems on the dispersion in metallicities of GC system as a function of host halo mass. Legend as in \autoref{fig:mean_mpb}.}
  \label{fig:std_mpb}
\end{figure}

\autoref{fig:mean_mpb} demonstrates the more drastic impact of satellite accretion on the MDF of GC systems. Without the contribution of accreted systems, the mean metallicity would continue to rise with halo mass, rather than plateauing at $\feh \approx -0.8$. The typical mean metallicity of GC systems in the most massive halos with $\Mh \approx 10^{14.5} \Msun$ would increase by about 0.5 dex relative to the observations and fiducial model. 

A related quantity is the dispersion in metallicities of a GC system, $\sigma_Z$ (the width of the MDF) plotted in \autoref{fig:std_mpb}. In \cite{choksi_etal_2018} we showed that the weak scaling of $\sigma_Z$ with $\Mh$ is a robust prediction of our model. However, when we exclude accreted GCs, we find a break in this relation at $\Mh \sim 10^{12.5}$, such that $\sigma_Z$ begins to \textit{decrease} with halo mass. In the most massive galaxies, the width of the metallicity distribution is far too narrow without the contribution of accreted satellite systems. In lower mass halos below $\Mh \approx 10^{12.5} \Msun$, $\sigma_Z$ still scales weakly with the halo mass because the main branch dominates cluster formation events.

\begin{figure}
\includegraphics[width=\columnwidth]{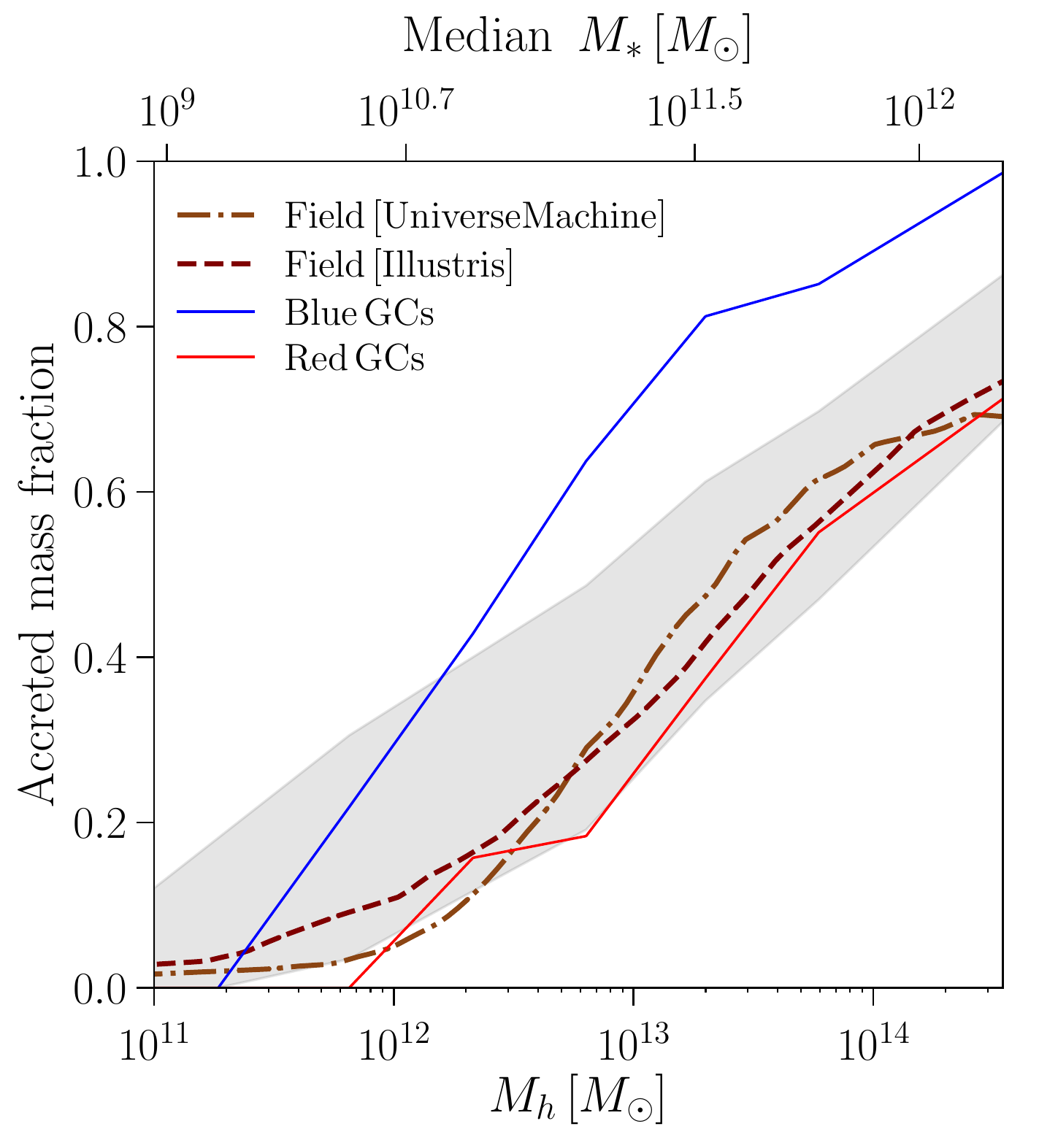}
\vspace{-5mm}
\caption{Fraction of mass accreted from satellites as a function of the host $z=0$ halo mass. Solid curves show the median model results for GCs, including splits for the red and blue GCs ($M_{\rm GC, red, sat}/M_{\rm GC, red}$ and $M_{\rm GC, blue, sat}/M_{\rm GC, blue}$). The brown dash-dot and maroon dashed curves show the result from the UniverseMachine \protect\citep{behroozi_etal_2019} and the prediction from the \textit{Illustris} simulation \protect\cite{rodriguez-gomez_etal_2016}, respectively.
} \label{fig:fsat}
\end{figure}

The results visualized in Figs.~\ref{fig:m_mpb}-\ref{fig:std_mpb} can be understood by analyzing \autoref{fig:fsat}. The shaded grey region in that figure shows the interquartile range for the accreted GC mass fraction for model halos. In low mass halos, in-situ GC formation dominates, while in giant halos the majority of GCs are formed ex-situ. This explains the increasing discrepancy between the ``All'' and ``MPB Only'' curves at high halo masses in the previous figures. Because satellites (and by extension, their GCs, according to the adopted galaxy mass-metallicity relation) will tend to have lower metallicities, excluding their contribution leads to a higher metallicity. 

We note that the fraction of clusters by \textit{number} formed in satellites follows a very similar trend for red clusters, but is always somewhat higher for blue clusters. This is because the mean mass of blue clusters formed in satellites is lower than that formed in the central galaxy by about a factor of 2. This is due to the fact that satellites have smaller gas reservoirs, which prevents sampling of the high-mass end of the cluster initial mass function (for a more detailed discussion of this effect see \citealt{choksi_gnedin_19}, Appendix A). For a Milky Way mass halo with $\Mh \approx 10^{12} \Msun$, we expect only $\approx$30\% of GCs to have formed ex-situ.

The figure also shows two estimates of the accreted stellar mass fraction for \textit{all} (``field'') stars. The dash-dot brown curve shows the semi-empirical result derived via forward modeling from the UniverseMachine model of \cite{behroozi_etal_2019}, while the dashed brown curve shows the prediction from the \textit{Illustris} cosmological simulation as analyzed by \cite{rodriguez-gomez_etal_2016}. These two curves can be compared to the shaded black region, showing the analogous result for GCs. All three curves show a similar normalization and slope for very massive galaxies, demonstrating that GCs indeed trace well the overall buildup of field stars in galaxies. However, in Milky Way-sized galaxies there remains a a factor of 2-3 difference in the median trend. Numerical values of the plotted fractions are given for reference in Table~\ref{tab:fsat}.

\autoref{fig:fsat} also shows the median accreted GC mass fraction separately for the red and blue populations: $M_{\rm GC,blue,sat}/M_{\rm GC,blue}$ and $M_{\rm GC,red,sat}/M_{\rm GC,red}$. Clusters are identified by applying the Gaussian Mixture Modeling (GMM) algorithm of \cite{muratov_gnedin10} to the MDF of each GC system. We take the metallicity where the magnitude of two Gaussians are equal as the dividing line between blue and red clusters. This metallicity is typically located in the range $\feh \approx -0.7$ to $-1.2$. Therefore, in systems with either too few clusters to reliably apply GMM or strongly unimodal MDFs, we continue to use a fixed cutoff $\feh = -1.0$ to differentiate red and blue clusters.

As expected, we find that the blue clusters are systematically more likely to form ex-situ than red clusters. However, for a Milky Way mass galaxy we still predict a majority  -- roughly 70\% -- of the mass in blue clusters to have formed in-situ. On the other hand, in the most massive galaxies we predict the majority of mass of \textit{red} clusters to have also formed ex-situ, due to the strong dependence of the accreted mass fraction with host halo mass.

\begin{figure}
\includegraphics[width=\columnwidth]{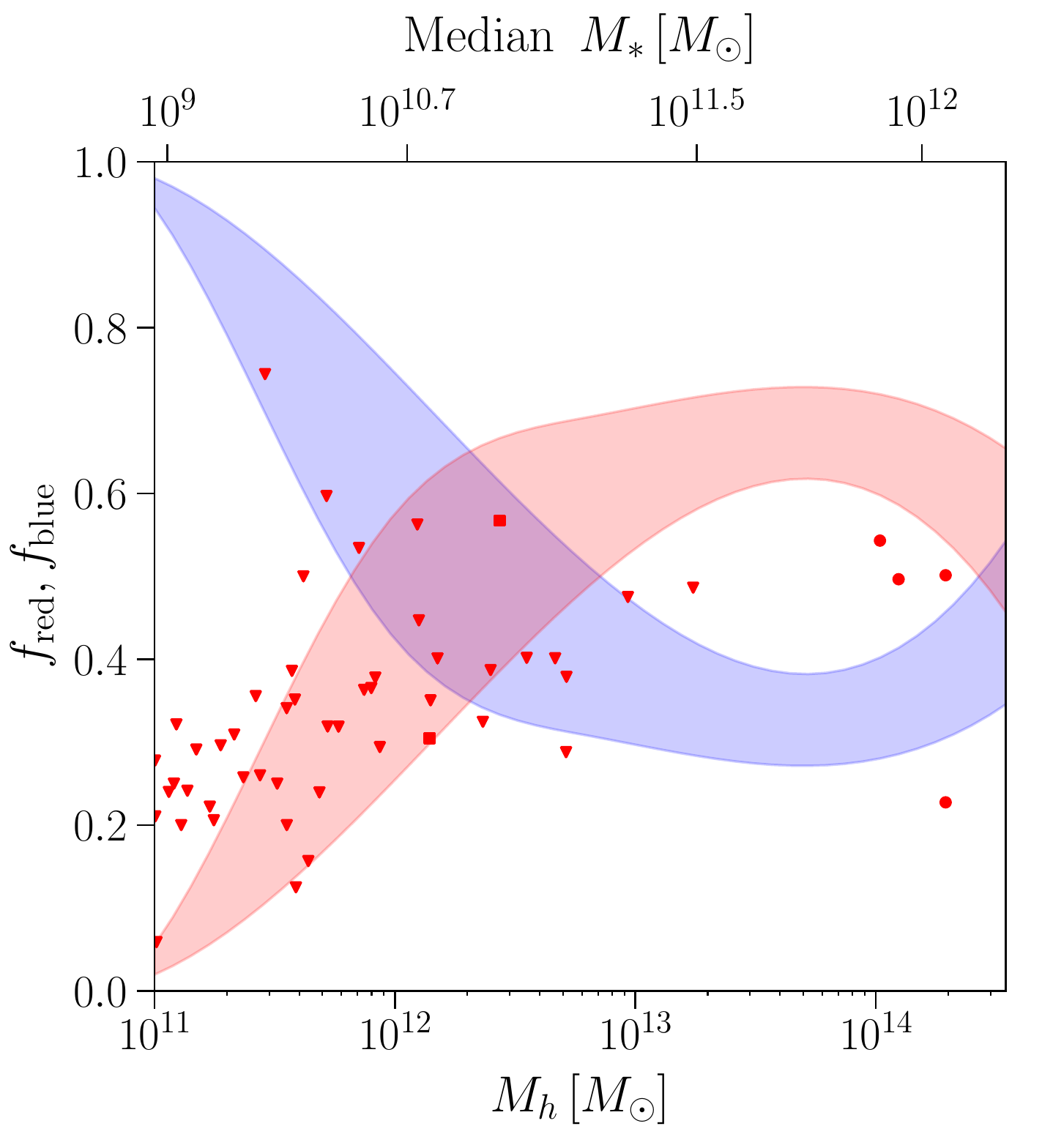}
\vspace{-5mm}
\caption{Fraction of clusters that are red and blue ($\fred + f_{\rm blue} = 1$). Shaded regions show the interquartile range for model galaxies in bins of halo mass. Upside-down triangles give upper-limits from the Virgo Cluster Survey \citep{peng_etal06}, the two squares represent the Milky Way and M31 \citep{harris96, huxor_etal_2014}, and circles represent brightest cluster galaxies \citep{harris_etal14_bcg1}. }
  \label{fig:fredblue}
\end{figure}

To better understand the relative importance of the red and blue populations as a function of galaxy mass, we show in Fig.~\ref{fig:fredblue} the fractions (by number): $\fred \equiv N_{\rm GC,red}/N_{\rm GC}$ and $f_{\rm blue} \equiv N_{\rm GC,blue}/N_{\rm GC}$. The model predicts $\fred$ to increase with halo mass. In dwarf galaxies with $\Mh \approx 10^{11} \Msun$, red clusters are highly subdominant, with $\fred \approx 10\%$. On the other hand, in group and cluster environments with $\Mh \gtrsim 10^{13} \Msun$, the model predicts more red than blue clusters, with $\fred$ reaching an asymptotic value of about 60\%. We note that \cite{el-badry_etal_2019} also predicted the fraction of red clusters as a function of halo mass (see their Fig. 9). It is reassuring that despite the many differences in their inputs, both models predict a very similar behaviour of $\fred$ with halo mass, including an asymptote at $\fred \approx 60\%$ for $\Mh \gtrsim 10^{13} \Msun$.

We have also overplotted in \autoref{fig:fredblue} upper limits (triangles) from the Virgo Cluster Survey \citep[VCS; ][]{peng_etal06} over the mass range $\Mh \approx 10^{11} - 10^{13} \Msun$. These points represent upper limits because HST covered only the central parts of their host galaxies, where red clusters preferentially reside. At $\Mh \gtrsim 10^{13} \Msun$, we also show data from the recent HST survey of brightest cluster galaxies \citep{harris_etal14_bcg1}. Finally, we include data from the Milky Way and M31 \citep{harris96, huxor_etal_2014}. The data are qualitatively consistent with the predictions of our model. For a Milky Way mass halo in particular, our model predicts $\fred \approx 30\%$, in excellent agreement with the observed fraction in the Galaxy \citep{harris96}.

\begin{table*}
\centering
\begin{tabular}{|c|c|c|c|c|c|}
\hline\\[-2mm]
Range of $\log_{10}\Mh/\Msun$ & $f_{\rm sat}$\,[IQR] & $f_{\rm sat, blue}$ & $f_{\rm sat, red}$ &  $f_{\rm sat, field}$ [Illustris] & $f_{\rm sat, field}$ [UniverseMachine]  \\
\hline
$11-11.5$  &  0.0  -  0.19  &  0.0  &  0.0  &  0.036  &  0.020  \\ 
$11.5-12$  &  0.035  -  0.31  &  0.17  &  0.0  &  0.091  &  0.034  \\ 
$12-12.5$  &  0.12  -  0.40  &  0.41  &  0.16  &  0.16  &  0.11  \\ 
$12.5-13$  &  0.19  -  0.49  &  0.62  &  0.16  &  0.28  &  0.29  \\ 
$13-13.5$  &  0.35  -  0.61  &  0.78  &  0.36  &  0.42  &  0.47  \\
$13.5-14$  &  0.47  -  0.70  &  0.86  &  0.56  &  0.56  &  0.62  \\ 
$14-14.5$  &  0.61  -  0.80  &  0.92  &  0.65  &  0.69  &  0.67  \\ \hline
\end{tabular}
\caption{Accreted mass fractions for several bins of halo mass, shown in \autoref{fig:fsat}. Column 1 gives the halo mass bin under consideration. Column 2 gives the interquartile range of the accreted mass fraction for all GCs. Columns 3 and 4 give the median accreted mass fractions for blue and red GCs, respectively. Columns 5 and 6 give the accreted stellar mass fractions for the field, from the \textit{Illustris} simulation \protect\citep{rodriguez-gomez_etal_2016} and the UniverseMachine \protect\citep{behroozi_etal_2019}. }
  \label{tab:fsat}
\end{table*}

\section{The role of galaxy mergers in triggering GC formation}
\label{sec:mergers}

In this section we analyze the role of mergers in our model. We emphasize that our criterion for forming clusters is only that the specific accretion rate onto a dark matter halo $R_m \equiv \dot{\Mh}/\Mh$ exceeds a threshold value $p_3$ which is calibrated based on observations; we do \textit{not} explicitly require any link to actual mergers of two distinct halos. Therefore, we can test what fraction of model clusters form in events that satisfy the criterion $R_m > p_3$ and also represent major mergers.

To identify mergers, for each halo in the merger tree we search for progenitors at the previous simulation snapshot. If more than one progenitor is identified, we label the event as a merger. However, the typical spacing between outputs of our adopted simulation is only $\approx$0.1~Gyr, yet the time for dark matter halos to merge (the interval between being accreted and disappearing from the halo catalog) can range from 0.1 to 1~Gyr. Therefore, the merger may be resolved over the course of multiple simulation snapshots. To account for the extended duration of mergers, we consider all the snapshots within some time interval $\Delta t$ \textit{before} the disappearance of the satellite from the halo catalog as ``undergoing'' a merger. The exact value of $\Delta t$ is somewhat arbitrary, as the range of merger timescales in the \textit{Illustris} simulation is quite broad, peaking at $\sim 200$~Myr but having an extended tail up to 1~Gyr. For clarity we adopt a fixed duration $\Delta t = 200$~Myr. A different choice of $\Delta t$ would affect the inferred merger probability (for example, $\Delta t = 400$~Myr increases it by a factor up to 1.6) but would not qualitatively alter our conclusions.

For each merger we calculate the ratio of masses of the satellite and central halo, $q < 1$. Because the merger may be resolved over multiple snapshots, the infalling satellite may have experienced significant tidal stripping before the last time it appears in the halo catalog. Therefore, the mass ratio calculated at the satellite's last snapshot is not representative of the extent to which the galactic potential was disturbed by the merger. The final recorded satellite mass is also strongly resolution-dependent; for a complete discussion of these effects, see \cite{rodriguez-gomez_etal_2015}. Instead, we calculate $q$ at the snapshot corresponding to the maximum mass of the satellite, which occurs approximately when the satellite crosses the virial radius of the central halo.\footnote{We verified our merger-identification process by calculating the cumulative number of major mergers along the main branch, in bins of $z=0$ halo mass, and compared this to the result in Fig. 7 of \cite{fakhouri_ma_2010} for the \textit{Millennium-II} simulation. We found generally good agreement in the redshift evolution and normalization of the number of mergers, with differences of $\lesssim 30\%$. The discrepancy is likely attributable to differences in the adopted collisionless simulation and details of the halo finding and merger identification process. Overall, this comparison gives us confidence in the number of major mergers we detect.}

Applying this procedure, we find that $\approx$20\% of all clusters with initial masses above $2 \times 10^5 \Msun$ form during major merger events with $q > \sfrac{1}{4}$, with only a weak scaling with the $z=0$ host halo mass.

\begin{figure}
\includegraphics[width=\columnwidth]{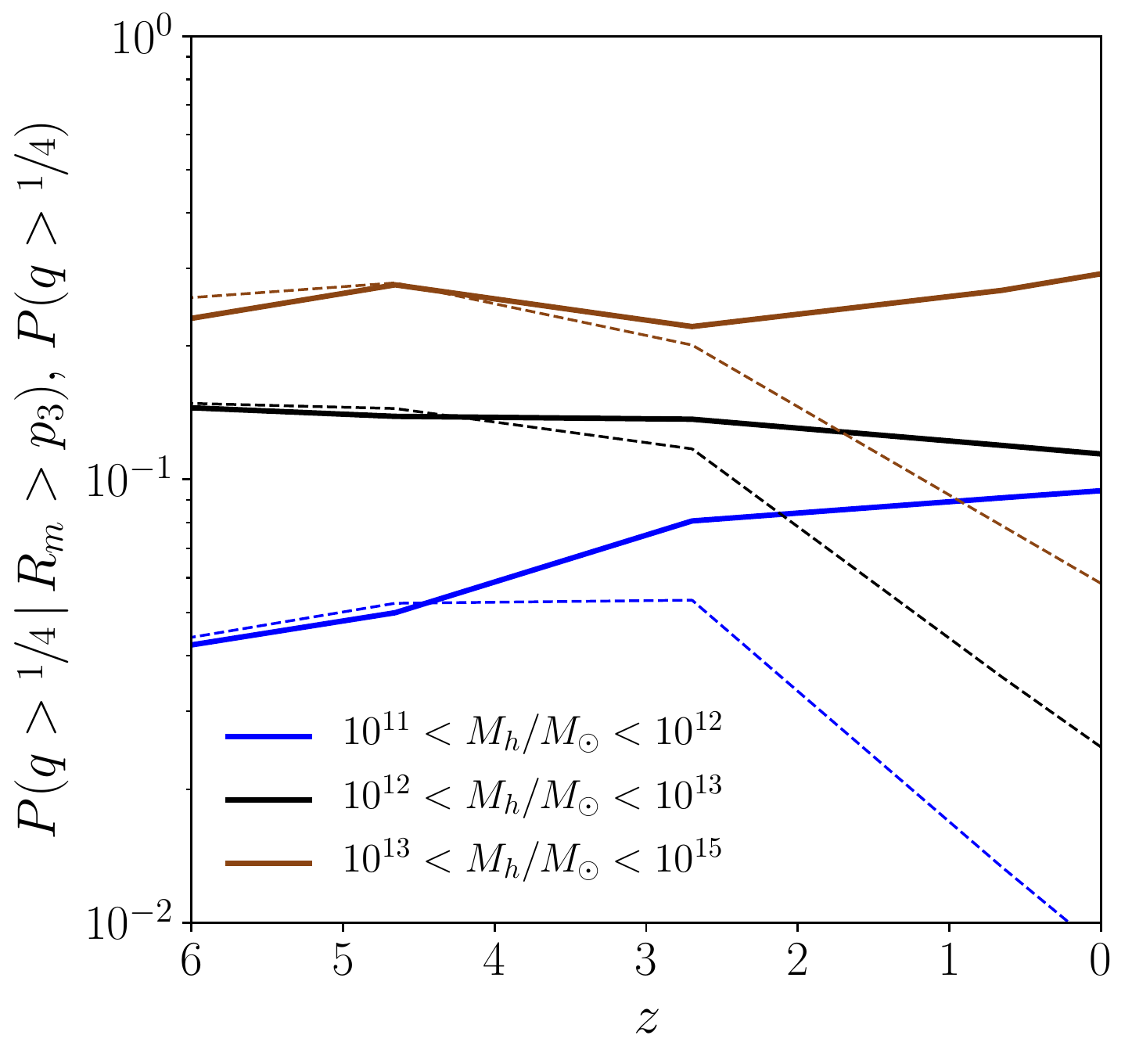}
\vspace{-5mm}
\caption{Solid lines show the conditional probability that a halo which is forming clusters ($R_m > p_3$) is \textit{also} undergoing a major merger ($q > 1/4$), while dashed lines show the probability that a halo is undergoing a major merger at a given redshift. We have adopted the fixed merger duration of 200~Myr before the disappearance of a satellite halo from the halo catalog (see text for details). Even though the merger rate decreases at low redshift, the importance of major mergers for triggering cluster formation remains significant because of the sharp decline in halo accretion rates.}
  \label{fig:dm_stats2}
\end{figure}

To better understand this result, we calculated at each simulation snapshot the ratio of the number of halos that satisfy both conditions $R_m > p_3$ and $q > \sfrac{1}{4}$ to the number of halos with $R_m > p_3$. In essence, this ratio represents the probability that a halo which is forming clusters is \textit{also} undergoing a major merger, $P(q > \sfrac{1}{4} \, | \, R_m > p_3)$. Our model assumes that the number of clusters formed in any event is proportional only to the cold gas mass in the central galaxy, and does not include the actual effects of the merger (increase of ISM pressure, rapid change of gravitational potential, etc.). Therefore, this probability can be taken as a simple estimate of the fraction of clusters expected to form during major mergers, which relies only on $\Lambda$CDM statistics and not on the details of our GC formation model. The solid curves in \autoref{fig:dm_stats2} demonstrate that this probability ranges from $\approx 5\%$ for halos with $\Mh \sim 10^{11} \Msun$ to $\approx 30\%$ for the highest-mass halos, with only a weak redshift evolution. On the other hand, the probability that a randomly selected halo is undergoing a major merger at a given redshift, $P(q > \sfrac{1}{4})$, actually \textit{decreases} steeply towards low redshift, as seen in the dashed curves in \autoref{fig:dm_stats2} which begin to diverge significantly from the solid curves at $z < 2.5$. 

\begin{figure}
\includegraphics[width=\columnwidth]{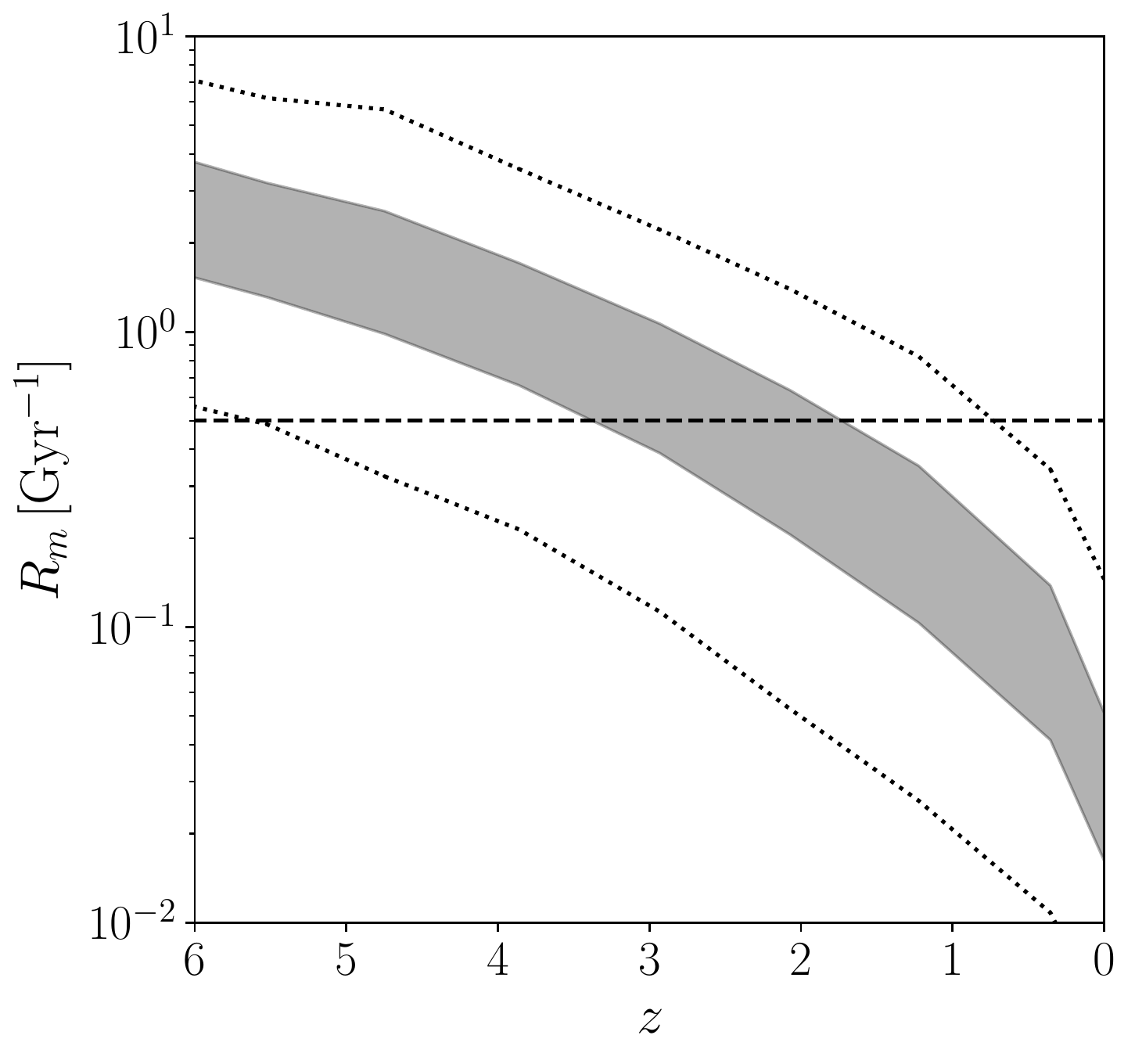}
\vspace{-5mm}
\caption{Evolution of the halo accretion rate for all halos in the \textit{Illustris-1-Dark} catalog. The shaded region shows the interquartile range (25th-75th percentiles) and the dotted lines give almost the full distribution (5th and 95th percentiles). The dashed line shows the threshold for forming clusters in our model of $\approx 0.5$~Gyr$^{-1}$.}
  \label{fig:dm_stats}
\end{figure}

Why does the probability of a halo undergoing a merger \textit{decrease} with time while the probability of major merger-induced cluster formation remains constant? This difference indicates that the probability of a halo exceeding the threshold for forming clusters $P(R_m > p_3)$ also systematically decreases with time. In terms of the probabilities it can be written as $P(q > \sfrac{1}{4}) = P(q > \sfrac{1}{4} \, | \, R_m > p_3) \, P(R_m > p_3)$. In \autoref{fig:dm_stats} we illustrate the sharp decline of the halo accretion rates $R_m$ with cosmic time, consistent with standard expectations from $\Lambda$CDM \citep[e.g.,][]{mcbride_etal_2009, behroozi_silk_2015, rodriguez-puebla_etal_2016}. At high redshift, $z > 3$, typical halo accretion rates significantly exceed the threshold $p_3$ and cluster formation can occur almost continuously. Of course for some halos occasional drops in $R_m$ below $p_3$ can occur (as demonstrated by the dotted 5th percentile curve) but these are rare. On the other hand, at $z \lesssim 1.5$ the entire region of the $R_m$ interquartile range lies below the cluster formation threshold. However, a few formation events do still occur (as demonstrated by the dotted 95th percentile curve). Thus low-redshift cluster formation proceeds far more discretely.

\section{Discussion}
\label{sec:discussion}

\subsection{The origin of the GC-halo mass relation}

\cite{el-badry_etal_2019} showed that the tight $\Mgc-\Mh$ relation does not \textit{necessarily} imply such a relationship at formation. The action of repeated galaxy mergers, which combine GC systems, naturally leads to a tight $\Mgc-\Mh$ relation by \mbox{$z=0$} simply due to the central limit theorem. However, this does not \textit{preclude} the existence of an $\Mgc-\Mh$ relation at formation; it only makes it unnecessary to explain local observations. 

As we showed in \autoref{sec:mgc_mh}, such a correlation is still expected at high redshift in our model, due to the adopted relation for the cluster formation rate \mbox{$\Mtot \propto M_g$}. This proportionality was initially motivated by the results of early cosmological simulations by \citet{kravtsov_gnedin05}. Those simulations identified dense giant gas clouds in high-redshift galaxies and applied a sub-grid model for the formation of globular clusters. The sum of masses of all clusters formed in a given event was roughly proportional to both the halo mass and the baryon (gas plus stars) mass. The relation to gas mass is more directly causal because star clusters are observed to form from dense cold gas. For this reason, the original version of our model \citep{muratov_gnedin10} adopted the proportionality of $\Mtot$ to the mass of neutral gas in the host galaxy. In \citet{choksi_etal_2018} we revised it to the mass of molecular gas, which should be even more closely related to the formation rate of massive clusters. While observations indicate that the surface density of molecular gas is most directly proportional to the star formation rate, our model cannot incorporate this relation because it does not calculate the spatial structure of galaxies. We believe our three-step scaling procedure from $\Mh$ to $\Mstar$ to $M_g$ to $\Mtot$ is an acceptable parameterization of the cluster formation rate.

\cite{kruijssen2015} predicted the $\Mgc-\Mh$ relation at $z=3$ to be nearly identical to the $z=0$ relation, in contrast to our prediction of a factor of $\sim$10 evolution (see \autoref{fig:mgcz3}). His model uses the cluster formation efficiency of \cite{kruijssen12} and assumes all GC formation happens in a single epoch. Furthermore, in contrast to our explanation of the shape of the $T_N - \Mstar$ relation, \cite{kruijssen2015} uses the shape of $T_N(\Mstar)$ to \textit{explain} the $\Mgc-\Mh$ relation and derive $\Mgc/\Mh \propto \Mstar^{0.1}$. However, in our model, it is the $\Mgc-\Mh$ relation that is more fundamental, because the mass that forms in clusters (indirectly) scales almost linearly with the halo mass.

Similar to our results, the fiducial model of \cite{el-badry_etal_2019} predicts a steeper-than-linear fall-off of the $\Mgc-\Mh$ relation at low galaxy mass. In their model, it is due to the cosmic UV background shutting off gas accretion onto dwarf galaxies. This effect is incorporated implicitly in our model in the shapes of the $\Mstar-\Mh$ and $\Mstar-M_g$ relations. In addition, our model includes the effect of systematically lower cluster masses in dwarf galaxies \citep{choksi_etal_2018}, which further steepens the relation. However, the scatter of the relation increases dramatically at the dwarf galaxy scale, which makes comparison with observations more challenging and requires a more detailed investigation.

\subsection{Accretion of satellite GC systems}

Most massive early-type galaxies in the local universe are believed to form in two phases: a short high-redshift phase of intense, concentrated in-situ star formation followed by late time accretion of satellite galaxies \citep[e.g.,][]{oser_etal10, forbes_etal_2011, van_dokkum_etal_2015}. Recently, \cite{beasley_etal_2018} studied the GC system of the nearby ``red nugget'' galaxy NGC 1277, so called because it is believed to be a low-redshift analog of early massive galaxies, i.e., it has not yet undergone the phase of late-time satellite accretion. They find NGC 1277 to be nearly devoid of blue GCs ($f_{\rm blue} \approx 17\%$) and present it as a puzzle for theory. This observation can be explained by our model, which predicts for typical galaxies of NGC 1277 mass ($\Mstar \approx 10^{11} \Msun$) that $\approx 40\%$ of all clusters are blue and $\approx 70\%$ of these blue clusters are accreted (Figs.~\ref{fig:fsat}-\ref{fig:fredblue}). Thus the blue fraction could drop to as low as 12\% if a galaxy happened to miss accretion of all satellites. This is plausible for NGC 1277 because it may have fallen early into the Perseus cluster and been tidally limited. In fact, the observed fraction $f_{\rm blue}$ indicates that some satellites did contribute to this GC system.

The predicted range of the fraction of blue clusters may thus be used to constrain the history of galaxy assembly. For the same stellar mass, galaxies in group or cluster environments where tidal limitation prevents accretion of satellite galaxies, should have a lower blue fraction than those in the field.

\subsection{Impact of galaxy mergers on the cluster formation rate}

\cite{li_etal_sim1, li_etal_sim2} used cosmological simulations of a single Milky Way mass halo to study in detail the formation process of massive star clusters. While our model ultimately produces similar results for the average GC formation epochs and demographics as in these simulations, the role of major mergers differs. \cite{li_gnedin19} find about 75\% of clusters with initial masses $M \geqslant 2 \times 10^5 \Msun$ form within 200~Myr of three major mergers, about a factor of 4 higher than our model prediction. The difference between these two results may stem from several sources. Galaxy formation simulations include a variety of detailed physics relevant for cluster formation that is not captured in our model. In particular, they find that the cluster initial mass function extends to higher masses during major mergers, increasing the probability of forming massive GCs that would survive to the present day. In contrast, our model adopts a merger-independent cluster formation rate. Finally, there is significant scatter in galaxy assembly histories, even at fixed galaxy mass, and therefore a single simulated realization may differ from the ensemble average of our model.

\subsection{Which model parameters disfavour mergers?}

Our current best-fit range on the free parameters $p_2$ and $p_3$ simultaneously matches properties of both the masses and metallicities of GC systems. Increasing the threshold accretion rate for forming clusters ($p_3$) and the cluster formation rate parameter ($p_2$) would lead to a higher fraction of all cluster formation occurring during major mergers. Such values could still match the observed $\Mgc-\Mh$ relation, by having fewer cluster formation events but more mass formed in each event. However, we found that such a parameter combination is disfavoured in our model. We have performed parameter optimization by minimizing the merit function \citep{choksi_etal_2018}: 
$$
  {\cal M} \equiv
  \frac{1}{N_h} \sum_h \left(\frac{\Mgc(z=0)}{\Mgcobs(\Mh)} - 1 \right)^2
  + \frac{1}{N_h}\sum_{h}  \left( \frac{0.58}{\sigma_{Z, h} } \right)^2
  + \frac{1}{G_{M}} + \frac{2}{G_{Z}},
  \label{eqn:merit}
$$
where the first term is the reduced $\chi^2$ of the $\Mgc-\Mh$ relation, the second term compares the observed average $\sigma_Z$ of 0.58 dex to the $\sigma_Z$ of model galaxies, and $G_M$ and $G_Z$ are the mass and metallicity ``goodness'' statistics, defined as the fraction of model-observed galaxy pairs that have GC system metallicity and mass distributions with an acceptable Kolmogorov-Smirnov statistic. 

To understand in detail what aspect of the model determines this conclusion, we investigated the relative contributions of each term in the merit function over a wide range $p_2 < 20$ and $p_3 < 3\,$Gyr$^{-1}$. We found that across almost the entire parameter space the variation of the $\chi^2$ and $G_Z$ terms dominates the gradient of the merit function. The $G_M$ and $\sigma_Z$ terms vary negligibly and could be eliminated for simplicity without loss of model accuracy.

High values of $p_2$ and $p_3$, which would lead to a higher fraction of cluster formation during major mergers, still result in a higher $\chi^2$ term than the fiducial model. At the same time, high values of $p_2$ are strongly disfavored by the $G_Z$ term: a higher normalization of the cluster formation rate would make the GC system MDF closer to the field star MDF, which is peaked at a significantly higher metallicity than the observed GCs.

Finally, we note that \autoref{fig:dm_stats} also demonstrates that the model results are not very sensitive to the \textit{exact} value of $p_3$. While shifting $p_3$ slightly may change results in detail, the median $R_m$ evolves by two orders of magnitude over cosmic time. Therefore, the main function of $p_3$ is instead to select out the general range of accretion rates $R_m$ at low redshift that \textit{are} allowed to form clusters.

\section{Summary}

In this work, we explored the origins and buildup of many of the GC system-host galaxy scaling relations. Our main results are:
\begin{enumerate}
    \item The ratio $\Mgc/\Mh \sim 10^{-4}-10^{-3}$ at birth, and evolves only mildly at $z > 3$ due to continuous cluster formation. From $z \approx 3$ to $z = 0$, the ratio decreases by roughly a factor of ten because the GC formation rate drops sharply while the halo mass continues to increase (\autoref{fig:mgcz3}).
    
    \item The shape of the relation between GC specific frequency $N_{\rm GC}/\Mstar$ and host galaxy stellar mass exhibits a characteristic $U-$shape, which is set by the near constancy of the ratio $\Mgc/\Mh$ and the peaked shape of the stellar mass-halo mass relation (\autoref{fig:tn}).
    
    \item The fraction of GC mass accreted from now-disrupted satellite galaxies increases monotonically with $\Mh$, ranging from a few percent at $\Mh \approx 10^{11} \Msun$ to 80\% at $\Mh \approx 10^{14.5} \Msun$. These values are similar to the accreted fraction of ``field'' stars in giant galaxies, but exceed them in Milky Way-sized and dwarf galaxies (\autoref{fig:fsat} and \autoref{tab:fsat}). Blue GCs are systematically more likely to form in satellite galaxies than red GCs.
    
    \item Consequently, without the contribution of accreted GCs the mean metallicities of GC systems would be up to 0.5 dex too high and the metallicity dispersions 0.4 dex too low (Figs.~\ref{fig:mean_mpb}-\ref{fig:std_mpb}). The combined mass in GCs would also be up to 0.3 dex too low (\autoref{fig:m_mpb}). 
    
    \item Major mergers are not the dominant channel for triggering cluster formation in our model. The extremely high accretion rates onto dark matter halos at high-redshift are sufficient to trigger cluster formation without the aid of mergers (\autoref{fig:dm_stats}). At lower redshift, major mergers become increasingly important, especially for giant galaxies (\autoref{fig:dm_stats2}).
    
    \item The fraction of clusters formed during major mergers in our model is a factor of 4 lower than in recent simulations of galaxy formation \citep{li_gnedin19}. This discrepancy may indicate a limitation of the analytical model in capturing detailed physics of GC formation. Nevertheless, both the analytical model and cosmological simulations predict similar distributions of GC formation epochs.
\end{enumerate}

\section*{Acknowledgements}
We thank Hui Li, Kareem El-Badry, Dan Weisz, Tom Zick, and Wren Suess for useful conversations and encouragement. We also thank the participants of the Lorentz center workshop ``Formation of Stars and Massive clusters in Dwarf Galaxies over Cosmic Time'' for useful discussions. This work made use of the \textsc{matplotlib} Python package \citep{hunter_etal_2007}. This work was supported in part by the National Science Foundation through grant 1412144.




\bibliographystyle{mnras}
\bibliography{GC3/gc_oleg,GC3/gc_nick} 






\bsp	
\label{lastpage}
\end{document}